\documentclass[runningheads]{llncs}
\usepackage{graphics}
\usepackage{epsfig}
\usepackage{graphicx}
\begin{document}
\title{Streamlined Data Fusion: Unleashing the Power of Linear Combination with Minimal Relevance Judgments}

\author{Qiuyu Xu, Yidong Huang, Shengli Wu and  Adrian Moore} 
\institute{a. School of Computer Science, Jiangsu University, Zhenjiang, China \\
       b. School of Computing, Ulster University, Belfast, UK}

\maketitle              
\begin{abstract}

Linear combination is a potent data fusion method in information retrieval tasks, thanks to its ability to adjust weights for diverse scenarios. However, achieving optimal weight training has traditionally required manual relevance judgments on a large percentage of documents, a labor-intensive and expensive process. In this study, we investigate the feasibility of obtaining near-optimal weights using a mere 20\%-50\% of relevant documents. Through experiments on four TREC datasets, we find that weights trained with multiple linear regression using this reduced set closely rival those obtained with TREC's official "qrels." Our findings unlock the potential for more efficient and affordable data fusion, empowering researchers and practitioners to reap its full benefits with significantly less effort.

\keywords{data fusion \and information retrieval \and linear combination \and weight training}

\end{abstract}

\section{Introduction}
Data fusion is a useful technology for various information retrieval tasks
to improve performance. Linear combination is a strong data fusion method.
If proper weights are assigned to component retrieval systems, then it is able to 
achieve better results than those methods such as CombSum \cite{Fox&Koushik93}, 
CombMNZ \cite{Fox&Koushik93}, Borda Count \cite{Aslam&Montague01}, 
which treat all component retrieval systems equally.

Weights assignment is a key issue for the success of linear combination.
Quite a few different weights assignment methods have been proposed 
\cite{Lillis&Toolan08,Lillis&Zhang10,Wu12-dke,Ghosh&Parui15,Sivaram&Batri20,Xu&Huang16}.
However, almost all of them are supervised learning methods
and a training data set is required. Usually, a training data set includes
a collection of documents $D$, a group of queries $Q$, 
a group of retrieval systems $IR$ and retrieval results $S$ from $IR$ corresponding to $Q$ and 
$C$, and relevance judgment $J$.
In many situations, to set up a proper training data set requires a lot of effort.
Among them, relevance judgment is an expensive component because it needs human judges to decide
which document is relevant to which query. It is especially so when the document collection is very large.
Probably this is one of the major reasons that very simple data fusion methods such as CombSum, CombMNZ, and Borda Count,
are more frequently used in various retrieval tasks \cite{Lillis20}. More complicated and expensive ones are rarely used,
although they can lead to better retrieval performance.

In this piece of work, we investigate if it is possible to train proper weights for linear combination 
using a ``lightweight'' training data set.
Rather than identify and use all relevant documents for each query, a ``lightweight'' one only includes
a subset of all the relevant documents for each query. 
Especially, we focus on the method of multiple linear regression
because it is very good for weights training.
Theoretically, the weights calculated by that method are optimum in the least squares sense \cite{Wu12-dke}.
Empirically, it outperforms many other weights assignment methods such as SlideFuse \cite{Lillis&Toolan08},
PosFuse \cite{Lillis&Zhang10}, MAPFuse \cite{Lillis&Zhang10}, and SegFuse \cite{Shokouhi07}. 
This is also confirmed later in this paper.

The remaining of this paper is organized as follows.
Section 2 reviews the method of weights assignment by multiple linear regression for linear combination.
Section 3 discusses relevance judgment, especially the pooling policy used in TREC. 
Section 4 presents the setting and experimental
results of this study. Some more analysis is given in Section 5. Finally, Section 6 makes some concluding remarks.

\section{Weights assignment by multiple linear regression}

A training data set comprises 
a collection of $l$ documents ($D$), a group of $m$ queries ($Q$), and 
a group of $n$ information retrieval systems ($IR$).
For each query $q^{i}$ $(1 \leq i \leq m)$, all information retrieval systems $ir_j$ $(1 \leq j \leq n)$
provide their estimated relevance scores to all the documents 
in the collection. Therefore, we have 
($s_{1k}^i$, $s_{2k}^i$,..., $s_{nk}^i$, $y_{k}^i$) 
for $i$ = (1, 2, ..., $m$), $k$ = (1, 2, ..., $l$).
Here $s_{jk}^i$ stands for the score assigned by retrieval system $ir_{j}$ 
to document $d_{k}$ for query $q^{i}$; $y_{k}^i$ is the judged relevance score of $d_{k}$
for query $q^{i}$. If binary relevance judgment is used, then it is 1 for relevant 
documents and 0 otherwise.

\( Y=\{y_{k}^i; i=(1,2,...,m), k=(1,2,...,l)\} \)
can be estimated by a linear combination of scores from all component systems.
Consider the following quantity 

\[ \mathcal{G}=\sum_{i=1}^m{\sum_{k=1}^l{{[y_{k}^i-
(\hat{\beta_{0}}+\hat{\beta_{1}}s_{1k}^i+\hat{\beta_{2}}s_{2k}^i+...+\hat{\beta_{n}}s_{nk}^i)]}^2}}  \]

\noindent when $\mathcal{G}$ reaches its minimum, the estimation is the most accurate. 
$\beta_{0}$, $\beta_{1}$,
$\beta_{2}$,..., and $\beta_{n}$, the multiple linear regression coefficients, are 
numerical constants that can be determined from observed data.

In the least squares sense the coefficients obtained by multiple linear regression can
bring us the optimum fusion results by the linear combination method,
since they can be used to make the most accurate estimation of the relevance scores 
of all the documents to all the queries as a whole \cite{Wu12-dke}.
$\beta_{j}$ can be used as weights for the fusion of retrieval systems $ir_j$ $(1 \leq j \leq n)$.

Ideally, \( Y=\{y_{k}^i; i=(1,2,...,m), k=(1,2,...,l)\} \) should include all relevant documents,
then it is possible to make the most accurate estimation.
If only partial relevant documents are identified, some measures should be taken to treat 
all component retrieval systems fairly. Thus possible bias towards one or a subgroup of retrieval systems
can be avoided. The pooling system used in TREC is a good practice and we also apply it in this study. 
See next section for more details.

\section{Relevance judgment}

For a given task in TREC, its organizer provides a relevance judgment file ``qrels''.
Usually this file is generated by applying a pooling policy.
That is, for all the runs submitted to that task or a carefully-selected subset of them,
a certain number of top-ranked documents are put into a pool.
All the documents in the pool are evaluated manually. All those documents that are not in the pool are
assumed as non-relevant. Such a pooling policy is cost effective and fair to all participants \cite{KeenanSK01}. 

The pooling policy can be divided into two types: fixed-length and variable-length \cite{LosadaPB19}.
In the fixed-length pooling with a given number $k$, the top-$k$ ranked documents
from all the runs are put into the pool for every query.
In the variable-length pooling, the number of documents taken into the pool may vary across queries \cite{Aslam&Pavlu06,CarteretteAS06,CormackG16}.
Comparing these two, the latter requires a little more effort but can be more effective. 
Up to now the fixed-length pooling is the most common methods used in most TREC events.
Therefore, we go with this policy and find it is good for our purpose, as discussed in the next section.
Having said that, It is still an interesting thing to see the effect of different pooling policies.
This remains to be our future work.

We used four data sets for this investigation. Two of them were
runs submitted to the TREC 2018 and 2019 precision medicine track \cite{RobertsDVHBL18,RobertsDVHBLPM19}
and the other two were runs submitted to the TREC 2020 and 2021 deep learning track \cite{CraswellM20,CraswellM21}. 
Table 1 shows the related information about them.
We observe that the TREC 2021 data set includes the most relevant documents, while the TREC 2020 data set
includes the least relevant documents.  The most documents were evaluated for the 2018 data set,
while the least documents were evaluated for the 2020 data set.

\begin{table}
\begin{center}
\caption{Information of the four TREC data sets (the literature articles task, precision medicine track in TREC 2018 and 2019, machine learning track in TREC 2020 and 2021) and their ``qrels'' files.}
\begin{tabular}{|c|c|c|c|c|c|}
\hline
 \multicolumn{3} {|c|} {Data set} & No. of  & No. of         & No. of        \\ \cline{1-3}
Year & Track & Task & Queries & Evaluated docs & Relevant docs \\  \hline
2018 & Precision medicine & Literature articles & 50 & 22429 & 5588  \\
2019 & Precision medicine & Literature articles & 40 & 18316 & 5544  \\
2020 & Deep learning & Passage retrieval & 54 & 11386 & 1666  \\
2021 & Deep learning & Document retrieval & 57 & 13058 & 8203  \\
\hline
\end{tabular}
\end{center}
\end{table}

In order to investigate the impact of poor length on the number of relevance ducuments included,
we generated a good number of relevant judgment files 
from the official qrels by the fixed-length pooling. More specifically, the procedure is as follows:
for all the runs in a data set, we look at top 2, 3,..., and up to 20 documents of the result lists
for each of the queries to see if they are labeled as relevant or not in the official qrels.
This is very similar to the pooling policy used in TREC apart from
one point: we do not make relevance judgment manually, but instead assuming that all relevant
documents have been identified in TREC's official qrels.
For each data set, we obtain 19 partial qrels and each of them includes a subset of relevant documents in the official qrels.

Fig. 1 shows the number of relevant documents included in these partial qrels,
while Fig. 2 shows the percentage of relevant documents in these partial qrels compared with the official full qrels.
Not surprisingly, both the number and the percentage increase with the pool length all the way through.
For the same data set, the shapes of the curves in both Fig. 1 and 2 are similar. 
One noticeable thing is the 2021 data set. It includes the largest number of relevant documents (8203 in total or 143.91 per query),
but the percentage of all relevant documents is the lowest.


 \begin{figure*}
 	\begin{minipage}[c]{0.4\linewidth}
 		\centering
 		\includegraphics[width=6cm]{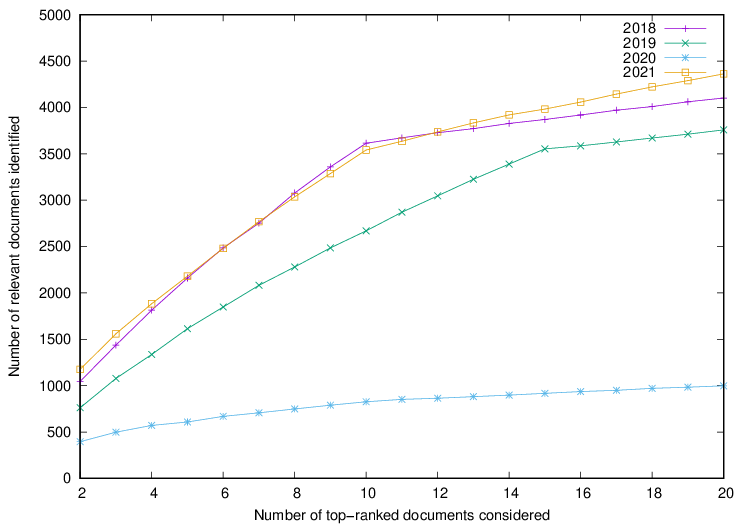}
 		\centering
 		\caption{Number of relevant documents identified with varying pool length}
 	\end{minipage} \hspace{1cm}
 	\begin{minipage}[c]{0.4\linewidth}
 		\centering
 		\includegraphics[width=6cm]{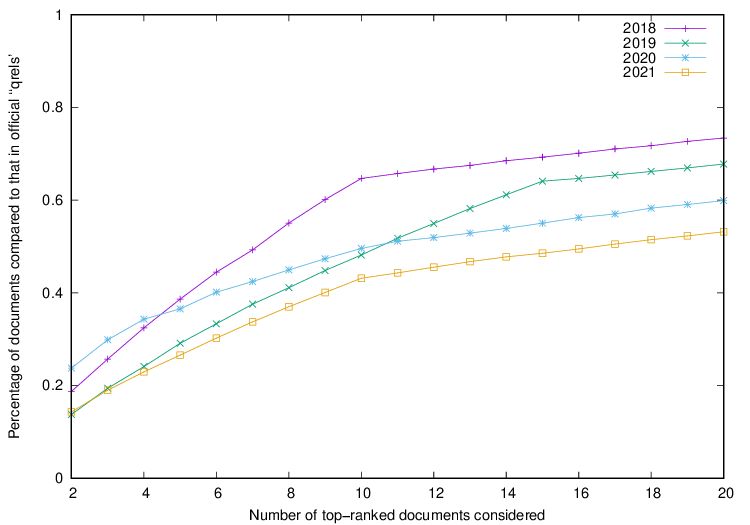}
 		\centering
 		\caption{Percentage of relevant documents identified with varying pool length}
 	\end{minipage}
\end{figure*}

\section{Experimental settings and results}

Four data sets, as described in Section 3, were used for this investigation.
In each data set, we selected a subset of runs for the fusion experiment. The selection criteria is:
we chose the best one in MAP from each participant, thus 19, 14, 15, and 16 runs were selected for each of the four data sets, respectively.
Such a selection policy would let the retrieval systems selected with more diversity,
because the multiple runs submitted by the same participant are retrieved from the same retrieval systems with
small difference on optional components, parameter settings, and so on, and they are
more similar than those submitted by different participants.
All the selected runs are listed in Appendix, in descending order of their MAP values. 

Apart from the official qrels, we used two partial qrels.
From all these 19 partial qrels, we choose two with roughly 20\% and 50\% of the relevant documents in the official qrels.
The pool lengths are 2 (18.74\%) and 6 (44.47\%) for the data set of 2018,
4 (24.10\%) and 10 (48.16\%) for the data set of 2019,
2 (23.77\%) and 10 (49.58\%) for the data set of 2020,
and 3 (19.02\%) and 17 (50.52\%) for the data set of 2021.
They are referred to as 20\% and 50\% partial qrels, respectively.

For all the runs in each year group, they were ranked by MAP values.
Then we fused top 2, 3, ..., until all of them by linear combination,
in which multiple linear regression was used for weights assignment. See Section 2 for details.
The two-fold cross-validation methodology was applied: 
for all the queries in a data set, 
we divided them into two partitions: odd-numbered and even-numbered. One partition was used for weights training and the other  
for testing, and vice versa. 
The reciprocal model, \(score(d)=1/(60+rank(d))\), 
was used to convert each document's ranking into a score because it is very good and reliable \cite{Cormack&Clarke09},
while raw scores from initial retrieval systems were not used.
Relevance judgment is required for weights training.
Apart from the official qrels file, we used two partial $qrels$ files, 
which consist of roughly 50\% and 20\% of the relevant documents in the official qrels file, respectively. 

Four metrics were used for evaluation: MAP, RP, P@10, and P@20.
MAP and RP are system-oriented metrics, while P@10 and P@20 are user-oriented metrics.

Fusion performance of linear combination, using three different qrels files for weights training, is shown in Figs. 3-6. 
We can see that fusion performance (measured by MAP) is very close for all of them, 
although in most cases the official qrels does a little better than the two partial qrels.
For all other three metrics not shown in Figs. 3-6, the situation is very similar.
A comprehensive comparison is summarized in Table 2.
In most cases, using a relevance judgment file with 20\% or 50\% of the
relevant documents identified, fusion performance is very close to 
that of using the official relevance judgment file ``qrels''.
The difference is below 3\% in all the cases.
In some situations, fusion performance of using partial qrels is even better than that of using official qrels.
For the 2018 data set and 50\% partial pool, the fusion results are even a little better than 
that of using the full official qrels in both P@10 and P@20.
The same happens to the 2021 data set with 50\% partial pool and measured in P@10.

\begin{figure*}
	\begin{minipage}[c]{0.4\linewidth}
		\centering
		\includegraphics[width=6cm]{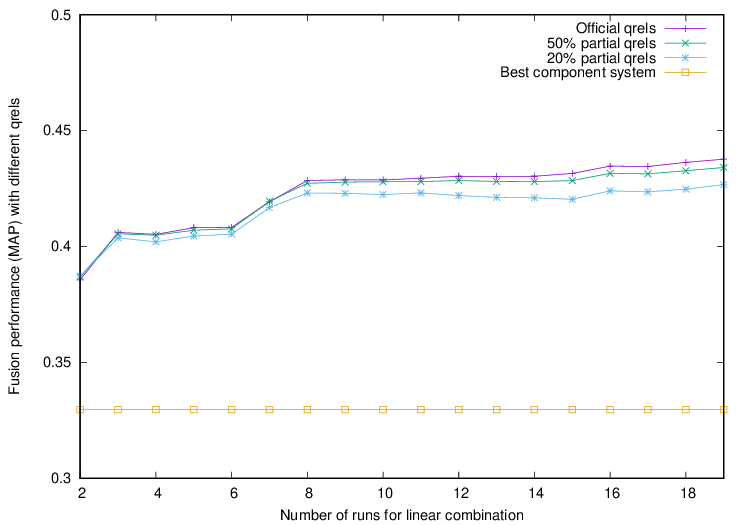}
		\centering
		\caption{Effect of three qrels (2018)}
	\end{minipage} \hspace{1cm}
	\begin{minipage}[c]{0.4\linewidth}
		\centering
		\includegraphics[width=6cm]{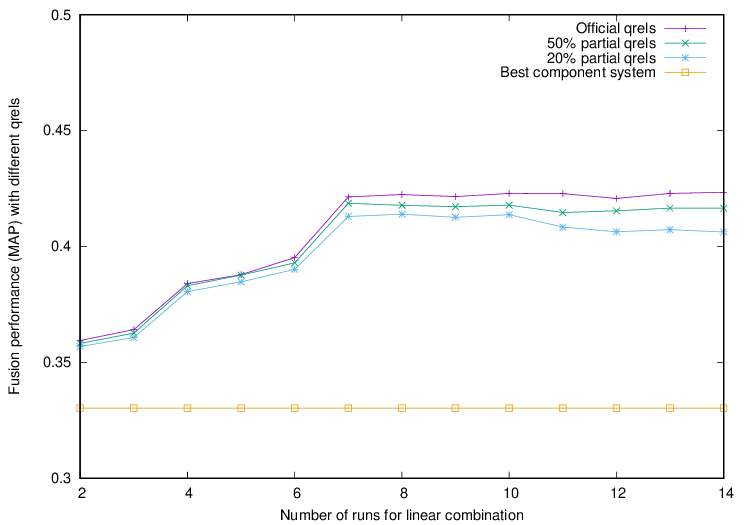}
		\centering
		\caption{Effect of three qrels (2019)}
	\end{minipage}
\end{figure*}

\begin{figure*}
	\begin{minipage}[c]{0.4\linewidth}
		\centering
		\includegraphics[width=6cm]{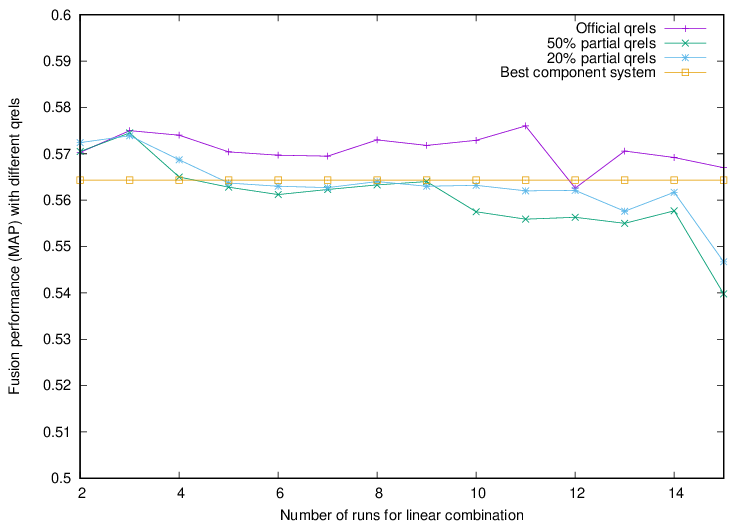}
		\centering
		\caption{Effect of three qrels (2020)}
	\end{minipage} \hspace{1cm}
	\begin{minipage}[c]{0.4\linewidth}
		\centering
		\includegraphics[width=6cm]{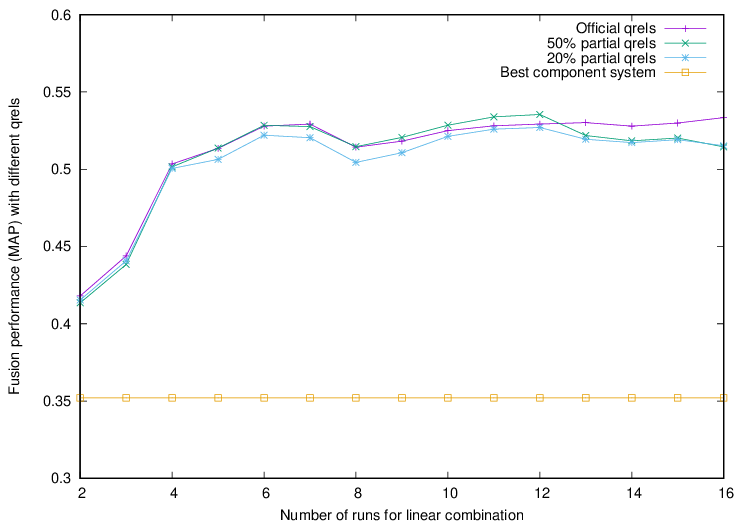}
		\centering
		\caption{Effect of three qrels (2021)}
	\end{minipage}
\end{figure*}

\begin{table} \label{deterioration}
\begin{center}
\caption{Performance comparison of linear combination using partial qrels versus official qrels for weights training}

\begin{tabular}{|l|rr|rr|rr|rr|} \hline

& \multicolumn{2}{|c|}{2018} & \multicolumn{2}{|c|}{2019} & \multicolumn{2}{|c|}{2020} & \multicolumn{2}{|c|}{2021} \\ 
\cline{2-9}
\raisebox{1.5ex}[0pt]{Metric} & 50\% & 20\% & 50\% & 20\% & 50\% & 20\% & 50\% & 20\% \\ \hline
MAP  & 99.62\% & 98.44\% & 99.06\% & 97.85\%   & 98.06\%  & 98.36\% & 99.45\% & 98.61\% \\ 
RP   & 99.86\% & 98.80\% & 98.62\% & 97.72\%   & 98.43\%  & 98.95\% & 99.73\% & 98.76\% \\
P@10 & 100.01\% & 97.97\% & 98.43\% & 98.05\%  & 98.70\%  & 99.27\% & 100.01\% & 99.96\% \\
P@20 & 100.03\% & 98.36\% & 99.35\% & 98.55\%  & 97.87\%  & 98.10\% & 99.61\% & 99.48\% \\ \hline

\end{tabular}
\end{center}
\end{table}

A few other data fusion methods including CombSum, CombMNZ, PosFuse, SlideFuse, 
and MAPFuse are also tested.
The results of two data sets 2018 and 2019 are shown in Figs. 7 and 8 for comparison. 
The results of other two data sets 2020 and 2021 are similar and not shown.
In both data sets, linear combination performs better than all 
four other fusion methods and the best component system on average. 
In Figs. 7 and 8, all data fusion methods do better than the best retrieval system involved with one exception.

\begin {figure*}
	\begin{minipage}[c]{0.4\linewidth}
		\centering
		\includegraphics[width=6cm]{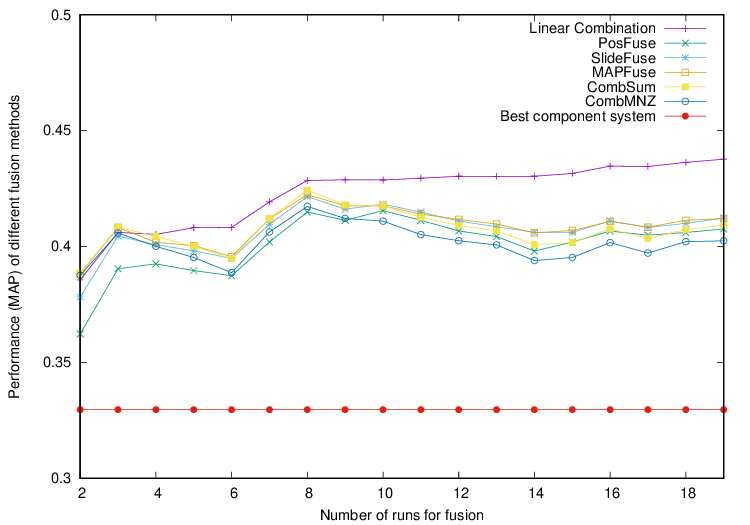}
		\centering
		\caption{Comparison of different fusion methods (2018)}
	\end{minipage} \hspace{1cm}
	\begin{minipage}[c]{0.4\linewidth}
		\centering
		\includegraphics[width=6cm]{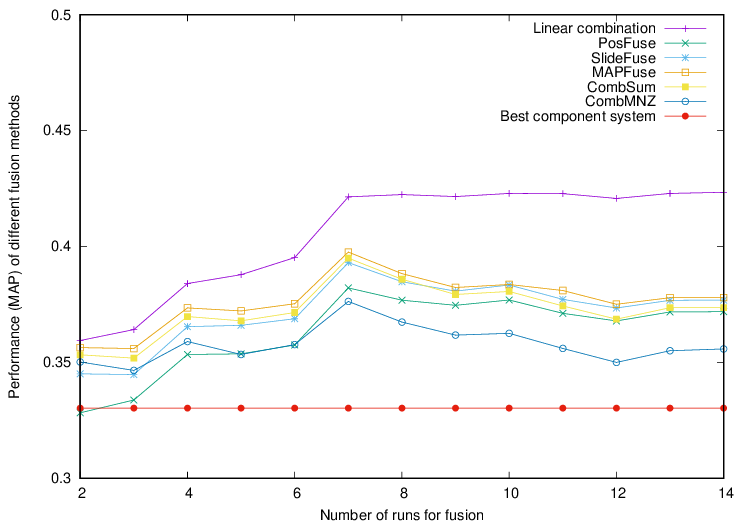}
		\centering
		\caption{Comparison of different fusion methods (2019)}
	\end{minipage}
\end{figure*}

\section{More observation and analysis}

When the pool length is very short, it may happen
that some queries only have very few relevant documents. 
It is interesting to find the effect of such queries to fusion results. 
The situation is more noticable for the 20\% partial qrels, therefore we take a look at it. 
For the four data sets 2018-2021, 8, 3, 42, and 4 queries include no more than 10 relevant documents, respectively.
Therefore, we report the results of two data sets 2018 and 2020.
See Table 3 for the average performances of all 50 queries, 
Group A (it includes queries with no more than 10
relevant documents), and Group B (it includes queries with more than 10 relevant documents).
We can see that Group B queries achieve better performance than Group A queries in all four metrics.
Especially for the 2018 data set, the performance of Group B  ``normal'' queries is roughly doubled 
compared with that of Group A queries with fewer relevant documents.

\begin{table} \label{less}
\begin{center}
\caption{The effect of queries with very few relevant documents in TREC 2018 and 2010 data sets (20\% partial qrels is used, 
Group A includes queries with no more than 10
relevant documents and Group B includes queries with more than 10 relevant documents)}

\begin{tabular}{|c|c|c|c|c|c|} \hline
Data Set & Query Set & MAP & RP & P\%10 & P\%20 \\ \hline
         & all 50 queries         & 0.4164 & 0.4375 & 0.6917 & 0.6433 \\ 
2018     & Group A (8 queies)     & 0.2317 & 0.2637 & 0.3569 & 0.3208 \\ 
         & Group B (42 queries)   & 0.4547 & 0.4741 & 0.7642 & 0.7119 \\ \hline
         & all 54 queries         & 0.5703 & 0.5441 & 0.6278 & 0.4704 \\ 
2020     & Group A (42 queries)   & 0.5131 & 0.4731 & 0.5214 & 0.3714 \\ 
         & Group B (12 queries)   & 0.6234 & 0.5888 & 0.8750 & 0.7417 \\ \hline

\end{tabular}
\end{center}
\end{table}

To see it in a more comprehensive way, we divide all the queries in a data set into three sub-groups  
of equal size based on the number of relevant documents identified.
We report the results of two data sets 2018 and 2019.
The three sub-groups include 17, 16, and 17 queries for the 2018 data set, and 13, 14, and 13
queries for the 2019 data set, respectively. ``Low'', ``Middle'', and ``High'' are used to name them.
For all those component systems and fused results by linear combination, their 
average performance for each of the sub-groups is presented in Table 4.
For both metrics P@10 \& P@20 and both average component results \& fusion results, Group ``High'' always obtains the highest value,
group ``Middle'' is in the middle, while group ``Low'' obtains the lowest value.
For MAP and RP, group ``Low'' always obtains the lowest values, while group ``Middle'' 
and ``High'' are competitive to be the winner.
It demonstrates there is a positive correlation between the number of relevant documents identified for 
given queries and retrieval performance for both component results
and fusion results.

\begin{table}
\begin{center}
\caption{The effect of number of relevant documents per query on component results and fusion performance 
(``No. R.'' stands for ``number of relevant documents''; F: fusion results; C: average performance of all 19 (2018) or 14 (2019) component results)}\label{ss}
\begin{tabular}{|c|c|c|c|c|c|}
\hline
 Group   & No. R. & MAP  & RP & P@10 & P@20  \\ \hline
Low (2018, F)      & 36.82  & 0.3766 & 0.3893 & 0.5471 & 0.4765  \\
Middle (2018, F)   & 94.94  & 0.5029 & 0.5150 & 0.7438 & 0.7157  \\
High (2018, F)    &  202.52 & 0.4311 & 0.4620 & 0.8750 & 0.8219  \\ \hline
Low (2018, C)      & 36.82  & 0.2022 & 0.2562 & 0.3706 & 0.3113  \\
Middle (2018, C)   & 94.94  & 0.2906 & 0.3447 & 0.5799 & 0.5266  \\
High (2018, C)    &  202.52 & 0.2445 & 0.3024 & 0.7418 & 0.6381  \\ \hline
Low (2019, F)     & 30.15  & 0.3233  & 0.3180 & 0.4231 & 0.3808  \\
Middle (2019, F)   & 100.93 & 0.4556 & 0.4696 & 0.6571 & 0.6571  \\
High (2019, F)    &  287.62 & 0.4887 & 0.4961 & 0.9154 & 0.8692 \\ \hline
Low (2019, C)     & 30.15  & 0.1848  & 0.2137 & 0.3198 & 0.2657  \\
Middle (2019, C)   & 100.93 & 0.2494 & 0.3000 & 0.5133 & 0.4834  \\
High (2019, C)    &  287.62 & 0.2376 & 0.3055 & 0.7390 & 0.6920 \\ \hline

\end{tabular}
\end{center}
\end{table}

Finally, we look at a related issue: when different qrels are used,
how does that affect various metric values?
For the same fusion results by linear combination 
and trained by using the official qrels, we calculate their MAP, RP, P@10, and P@20 values by
using the official and two partial qrels, respectively.
Table 5 shows the values of two data sets 2018 and 2019.
When using the two partial qrels, the variances of fusion performance from that of using the official qrels are also given.  
We can see that MAP and RP are less affected than P@10
and P@20. Because MAP and RP values are not directly linked to the number of relevant documents
in the whole collection, they are more robust and insensitive to the changes
in qrels.

\begin{table}
\begin{center}
\caption{The effect of partial qrels on fusion performance evaluation (the figures in parentheses indicate performance variances between partial qrels and official full qrels).}\label{qrels}
\begin{tabular}{|c|r|c|c|c|c|}
\hline
 Data set & qrels    & MAP & RP & P@10 & P@20 \\
\hline
      & full & 0.4230 & 0.4428 & 0.7064 & 0.6540 \\
TREC      &          & 0.4366 & 0.4450 & 0.6854 & 0.6088 \\
    & \raisebox{1.5ex}[0pt]{50\%}         & (+3.22\%) & (+0.50\%)  & (-2.97\%) & (-6.91\%) \\ 
2018          &      & 0.4151 & 0.4143 & 0.5751 & 0.4458 \\
          & \raisebox{1.5ex}[0pt]{20\%}         & (-1.87\%) & (-6.44\%) & (-18.59\%) & (-31.83\%) \\ 
\hline
      & full & 0.4225 & 0.4307 & 0.6695 & 0.6345 \\
TREC      &          & 0.4345 & 0.4443 & 0.6606 & 0.5991 \\
 & \raisebox{1.5ex}[0pt]{50\%} & (+2.8\%) & (+4.2\%) & (-1.3\%) & (-5.5\%) \\ 
2019      &      & 0.3922 & 0.3997 & 0.5767 & 0.4691 \\
          & \raisebox{1.5ex}[0pt]{20\%}         & (-7.2\%) & (-7.2\%) & (-13.9\%) & (-26.1\%) \\ 
\hline
\end{tabular}
\end{center}
\end{table}

Let us consider an example. If there are two relevant documents in the collection for 
a given query. One result list includes both, and one is at rank 1 and the other at rank 20,
then its MAP is (1+2/20)/2=0.55. If the one at rank 20 is regarded as non-relevant
in a partial qrels, then its MAP becomes 1! This can explain why in both 
data sets, both MAP and RP values with the 50\% partial qrels are even slightly higher than their counterparts 
with the official qrels. Such a situation will never happen to P@10 and P@20. 
If some relevant documents are removed from a qrels file, 
we will obtain equal or lower P@10 and P@20 values for the same result list. 
In the official qrels of TREC 2018, only three queries have less than 10 relevant documents
and all the others have more than 20 relevant documents; 
while in its 20\% partial qrels, eight queries have no more than 10 relevant 
documents, and 12 more queries have no more than 20 relevant documents, such a big difference 
can explain why P@20 decreases by (0.6540-0.4458)/0.6540=31.83\%.

\section{Conclusions}

For linear combination in data fusion, usually supervised methods are used for weights training. 
In this paper we have presented a method with partial relevance judgment.
Through four data sets from TREC, we have demonstrated that
using a small percentage of the relevant documents, 
the trained weights by multiple linear regression are almost as good as using all the
relevant documents in TREC's official qrels.
This finding is very helpful for the data fusion technology 
to be used in a more affordable way, and enable researchers and practitioners to get the full
benefit of it but with much less effort.

As our future work, we would consider extensions in two directions. One is to consider other alternative
methods to the simple fixed-length pooling. Although the fixed-length pooling works well in this study, 
it is interesting to find how other methods such as variable-length pooling \cite{LosadaPB19,Aslam&Pavlu06,CarteretteAS06,CormackG16}
perform for this purpose. 

Also related to this work, another research issue is to investigate 
the effect of partial qrels on some optimization-based weights training methods 
for linear combination such as the genetic algorithms \cite{Ghosh&Parui15,Sivaram&Batri20}
and differential evolution \cite{Xu&Huang16}. 

%
%

\newpage

\section{Appendix}

\begin{table}
\begin{center}
\caption{The runs selected in each of the four data sets for fusion experiment}
\begin{tabular}{|c|r|l|l|r|l|l|r|l|l|}
\hline
D & N & Run selected & MAP & N & Run selected & MAP & N & Run selected & MAP \\
\hline
2 & 1 & hpipubboost        & 0.330  & 2  & SIBTMlit4         & 0.322  & 3   & UCASSA2           & 0.320 \\
0 & 4 & imi\_mug\_abs2     & 0.318  & 5  & MedIER\_sa13      & 0.317  & 6   & UDInfoPMSA1       & 0.283 \\
1 & 7 & MSIIP\_PBPK        & 0.270  &  8 & UTDHLTRI\_NL      & 0.259  & 9   & SINAI\_Base       & 0.258 \\
8 & 10 & RSA\_DSC\_LA\_4   & 0.253  & 11 & KL18absHY         & 0.230  & 12  & cbnuSA1           & 0.228 \\
  & 13 & two\_stage	   & 0.228  & 14 & method\_fu        & 0.225  & 15  & mayomedsimp       & 0.199 \\
  & 16 & PM\_IBI\_run1     & 0.172  & 17 & cubicmnzAbs       & 0.161  & 18  & aCSIROmedAll      & 0.144 \\
  & 19 & bool51            & 0.138  &    & Average           & 0.245  & &   & \\ \hline
2 & 1  & jlpmcommon2       & 0.330 & 2   & DutirRun2         & 0.281  & 3   & imi\_mug2         & 0.276 \\
0 & 4  & SIBTMlit5         & 0.266 & 5   & SAsimpleLGD       & 0.260  & 6   & BM25              & 0.258 \\  
1 & 7  & ccnl\_sa2         & 0.254 & 8   & bm25\_680         & 0.250  & 9   & MedIR3            & 0.249 \\
9 & 10 & sils\_run2        & 0.224 & 11  & sa\_ft\_rr        & 0.188  & 12  & cbnuSA2           & 0.186 \\ 
  & 13 & absrun1           & 0.119 & 14  & default1m         & 0.004  &     & Average           & 0.225 \\   \hline
2 & 1  & p\_d2q\_rm3\_duo  & 0.564 & 2   & pash\_f3          & 0.550  & 3   & CoRT-electra      & 0.538 \\   
0 & 4  & NLE\_pr3          & 0.523 & 5   & RMIT-Bart         & 0.512  & 6   & bigIR-T5-BERT-F   & 0.510 \\  
2 & 7  & fr\_pass\_roberta & 0.499 & 8   & pinganNLP3        & 0.492  & 9   & 1                 & 0.490\\
0 & 10 & bcai\_bertl\_pass & 0.464 & 11  & nlm-ens-bst-2     & 0.460  & 12  & relemb\_mlm\_0\_2 & 0.435 \\ 
  & 13 & TUW-TK-2Layer     & 0.418 & 14  & bl\_bcai\_mdl1\_vt & 0.338 & 15  & terrier-InL2      & 0.313 \\
  &   & Average            & 0.474 &     &           & & & & \\   \hline
2 & 1 & pash\_doc\_f5      & 0.352 & 2   & NLE\_D\_v1        & 0.313  & 3   & uogTrDDQt5        & 0.296 \\
0 & 4 & watdff             & 0.296 & 5   & bigram\_qe\_cedr  & 0.296  & 6   & d\_f10\_mt53b     & 0.293 \\  
2 & 7 & doc\_full\_100     & 0.289 &  8  & uogTrBaseDDQC     & 0.288  & 9   & Fast\_Forward\_2  & 0.278 \\
1 & 10 & parade\_h3        & 0.259 & 11  & bcai\_bertm1\_ens & 0.256  & 12  & ielab-AD-uni-d    & 0.138 \\ 
  & 13 & CIP\_run2         & 0.248 & 14  & TUW\_IDCM\_S4     & 0.246  & 15  & max-firstp-pass   & 0.232 \\
  & 16 & webis-dl-3        & 0.225 &     & Average & 0.276 & & & \\ \hline
\end{tabular}
\end{center}
\end{table}

\end{document}